\documentclass[%
 reprint,
 amsmath,amssymb,
 aps,
floatfix,
]{revtex4-2}

\usepackage{graphicx}
\usepackage{dcolumn}
\usepackage{bm}
\usepackage{float}
\usepackage[hidelinks]{hyperref}
\usepackage{color}
\usepackage{upgreek}
\usepackage{comment}
\usepackage{ulem}

\newcommand{\SiN}{$\mathrm{Si_3N_4}$}
\newcommand{\um}{$\upmu$m}
\newcommand{\m}[1]{$\mathrm{#1}$}

\begin{document}

\preprint{APS/123-QED}

\title{Broadband Thermal Noise Correlations Induced by Measurement Back-Action}

\author{Jiaxing Ma}
\author{Thomas J. Clark}
\author{Vincent Dumont}
 \altaffiliation[Now at ]{Laboratory for Solid State Physics, ETH Zurich, CH-8093 Zurich, Switzerland}
\author{Jack C. Sankey}
 \email{jack.sankey@mcgill.ca}
\affiliation{%
 Department of Physics, McGill University, Montreal, Quebec H3A 2T8, Canada
}%

\date{\today}

\begin{abstract}

Modern mechanical sensors increasingly measure motion with precision sufficient to resolve the fundamental thermal noise floor over a broad band. Compared to traditional sensors -- achieving this limit only near resonance -- this capability provides massive gains in acquisition rates along with access to otherwise obscured transient signals. However, these stronger measurements of motion are naturally accompanied by increased back-action.
Here we show how resolving the broadband thermal noise spectrum reveals back-action-induced correlations in the noise from many mechanical modes, even those well-separated in frequency. As a result, the observed spectra can deviate significantly from predictions of the usual single-mode and (uncorrelated) multimode models over the broad band, notably even at the mechanical resonance peaks. This highlights that these effects must be considered in all systems exhibiting measurement back-action, regardless of whether the resonances are spectrally isolated or the readout noise is high enough that the noise peaks appear consistent with simpler models.
Additionally, these correlations advantageously allow the thermal noise spectrum to reach a minimum -- equivalent to that of a single mode -- in a band far from the resonance peak, where the mechanical susceptibility is comparatively stable against frequency noise.
\end{abstract}
\maketitle

\noindent

The most sensitive mechanical measurements have traditionally relied on resonance to enhance signals above readout noise, in order to achieve thermally limited operation \cite{saulson1990thermal,bachtold2022mesoscopic} in a narrow band near the mechanical frequency. As readout techniques continue to improve, it is now increasingly common to achieve this fundamental noise limit far from resonance \cite{numata2003wide,kajima1999wide}, even with modern low-noise mechanical elements \cite{fedorov2018evidence, Cripe2019Apr,zhou2021broadband, Cripe2019Apr}. This broadband regime promises massive gains in the rate at which information is gathered, while also unlocking sensitivity to transient (and other broadband) signals that would have otherwise been obscured.

Achieving this regime with modern sensors requires stronger measurement, meaning back-action (e.g., radiation pressure) necessarily plays a larger role. Usually this first appears as optical spring and damping effects \cite{Cuthbertson1996Jul, Sheard2004May, Arcizet2006Nov, aspelmeyer2014cavity}, but one can also observe quantum fluctuations in the back-action forces \cite{purdy2013observation,teufel2016overwhelming,peterson2016laser}, even now over a broad band \cite{Cripe2019Apr}.

Here we show how this back-action serves to correlate the noise from \textit{many} mechanical modes in a way that strongly impacts the observed thermal noise spectra over a broad band. Similar narrowband correlations have been observed near two degenerate resonances \cite{de2022coherent} and, by resolving thermal noise over the broad band, we reveal striking features involving a large number of modes -- even those that are well-separated in frequency. Equally importantly, we find these effects are large even at the observed resonance peaks, where one might expect the resonantly enhanced response to dominate over the broad and comparatively small background from other modes. Because of this, these effects must be considered in all systems exhibiting such back-action effects, even in the presence of high readout noise, where well-isolated resonance peaks appear to behave consistently with traditional single- or multi-mode models. Finally, as a practical consideration, we find that the lowest force noise -- approaching that of a single mode -- advantageously always occurs near the \textit{bare} mechanical frequency, allowing one to exploit the optical spring to shift unstable resonance peaks away from this optimal measurement band.

\begin{figure}[h]
\includegraphics[width=\columnwidth]{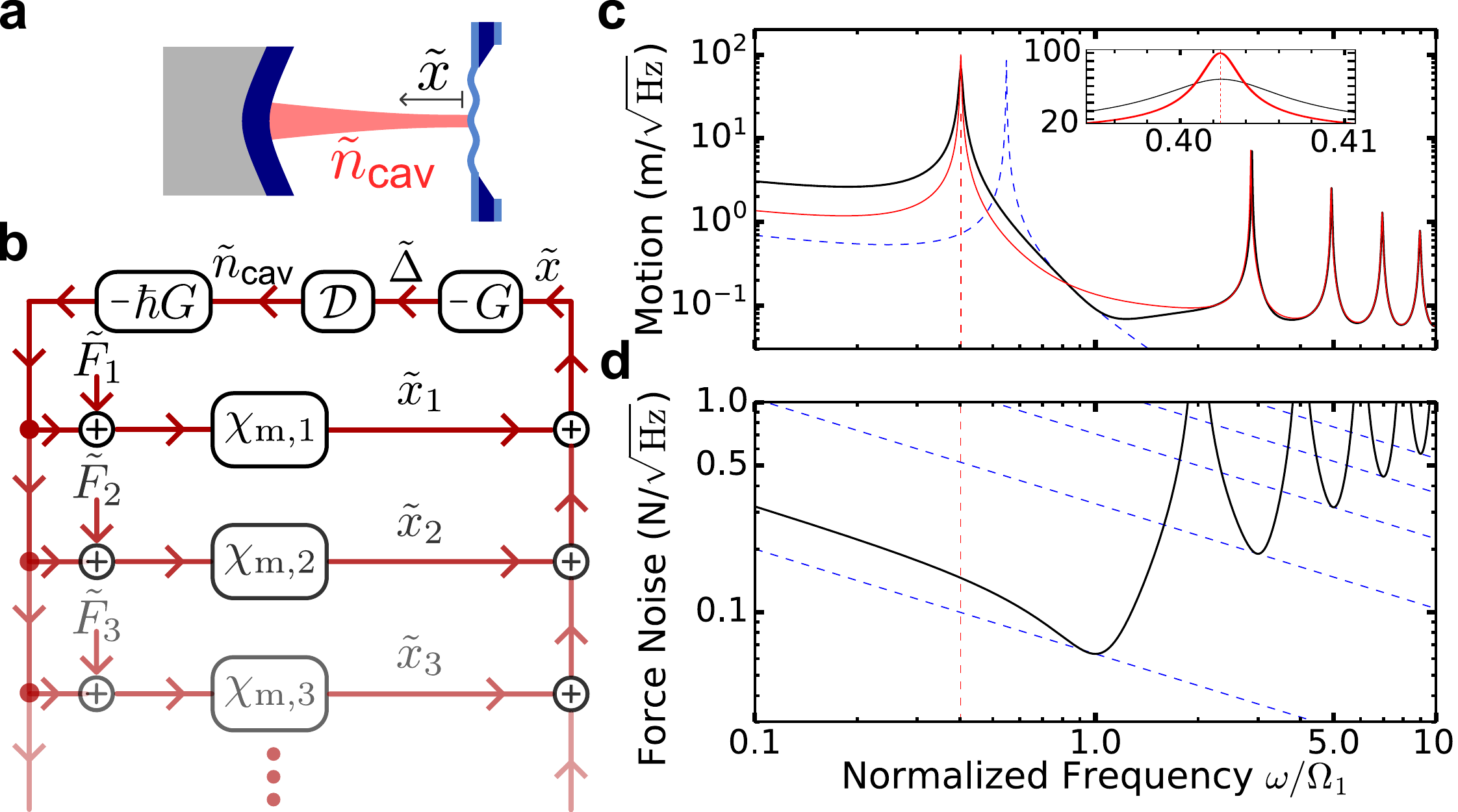}
\caption{\label{fig:1} Model for back-action-induced thermal noise correlations. (a) Canonical optomechanical system with a multimode mechanical system for one mirror. 
(b) Feedback loop describing (a), with one row per mechanical mode, each converting forces (including uncorrelated thermal noise $\tilde{F}_j$) to displacements $x_j$ through mechanical susceptibilities $\chi_{\text{m},j}$. The optomechanical coupling $G$ converts the total $\tilde{x}$ to detuning $\tilde{\Delta}$, and the cavity's detuning susceptibility $\mathcal{D}$ (just a constant in the fast-cavity limit) converts this to photon number $\tilde{n}_\text{cav}$, which finally applies radiation force $-\hbar G n_\text{cav} $ to all modes simultaneously, correlating displacements $\tilde{x}_j$.
(c) Plot of expected thermal displacement noise amplitude spectrum $\sqrt{S_x}$ including $N$=1,001 modes (black curve) of an ideal string with a cavity mode positioned at its center (frequencies $\omega$ normalized by $\Omega_1$). For this example, the cavity is driven by a red-detuned laser, applying anti-spring and shifting all modes to lower frequency. The blue dash curve includes only the fundamental ($j=1$), which exhibits less shift. The red curve shows the result treating all $N$ modes independently, but increasing $G$ to match the fundamental resonance frequency, highlighting major broadband and (inset) resonant discrepancies. (d) Dividing $\sqrt{S_x}$ by the total susceptibility to external forces (at the cavity spot) yields a force noise floor $\sqrt{S_F}$ (black). The dashed blue lines represent the (structural) force noise \cite{saulson1990thermal} for each individual mode. Importantly, the noise at the shifted resonance frequency (red dashed line) is significantly different than expected for a single mode. Moreover, the single-mode limit is approached at the \textit{bare} mechanical frequency, far from the modified resonance, where the spectrum is comparatively insensitive to frequency noise artifacts.
}
\end{figure}

\textit{Model.---}
To see how back-action induced noise correlation (BANC) presents itself, consider the optomechanical system drawn in Fig.~\ref{fig:1}(a), comprising a Fabry-Perot optical cavity in which one mirror is a flexible object with many mechanical modes (e.g., a membrane, string, ribbon, or cantilever), the combined motion of which modulates cavity resonance frequency $\omega_c$ to create dispersive coupling $G=\partial_x \omega_c$ \cite{aspelmeyer2014cavity}. This can be modeled by the feedback loop drawn in Fig.~\ref{fig:1}(b), with mechanical modes (indexed by $j$) independently driven by uncorrelated environmental (e.g., thermal) force noises of amplitude $\tilde{F}_{j}$ at frequency $\omega$. With structural damping \cite{saulson1990thermal}, a mode with inertial mass $m_j$, bare resonance frequency $\Omega_j$, and quality factor $Q_j$ has susceptibility
\begin{equation}
    \chi_{\text{m},j} = \frac{1/m_{j}}{\Omega_{j}^{2}-\omega^{2}-i\Omega_{j}^2/Q_{j}}
\end{equation}
that nominally converts $\tilde{F}_j$ to uncorrelated displacement $\tilde{x}_j=\chi_{\text{m},j}\tilde{F}_{j}$ (about equilibrium value $\bar{x}_j$), but with the cavity driven to mean photon occupancy $\bar{n}$ at mean detuning $\bar{\Delta}$ from resonance, the optomechanical coupling $G$ converts the summed displacement $\tilde{x} =\sum_j \tilde{x}_j$ to detuning fluctuations $\tilde{\Delta}=-G\tilde{x}$ that modulate the cavity photon number $\tilde{n}=-\mathcal{D}\tilde{\Delta}$ according to the optical cavity's detuning susceptibility $\mathcal{D}(\bar{\Delta},\bar{n})=\partial_\Delta n$ \cite{aspelmeyer2014cavity,clark2025detuning} (just a constant in the fast-cavity limit $\Omega_\text{m}\ll \kappa$, where $\kappa$ is the cavity's energy decay rate). Finally, the optomechanical coupling factor $\hbar G$ converts this fluctuation back to a radiation force applied to \textit{all} modes in parallel, closing the loop and causing BANC.

Combining the aforementioned transfer functions we arrive at a self-consistent linear system of equations, with the displacement of each mode $l$ satisfying
\begin{equation}
    \tilde{x}_{l} =\chi_{\text{m,}l}\tilde{F}_{l}-\xi\chi_{\text{m,}l}\sum_{j}\tilde{x}_{j},
\end{equation}
where $\xi\equiv \hbar G^{2}\mathcal{D}$ represents the strength of the back-action; in the fast cavity limit, this is just the optical spring constant. Solving this system for $\tilde{x}_l$ (using the Sherman-Morrison equation) yields
\begin{equation}
    \tilde{x}_{l}=\chi_{\text{m,}l}\left[\tilde{F}_{l}-\eta\xi\sum_{j}\chi_{\text{m,}j}\tilde{F}_{j}\right]
\end{equation}
with optomechanical ``interference factor''
\begin{equation}
    \eta\equiv\frac{1}{1+\xi\sum_{k}\chi_{\text{m,}k}},
\end{equation}
such that the \textit{total} displacement
\begin{equation}\label{eq:total-displacement-noise}
    \tilde{x} = \eta\sum_{j}\chi_{\text{m,}j}\tilde{F}_{j}.
\end{equation}
If the driving noises $\tilde{F}_j$ are uncorrelated and characterized by power spectral densities $S_{F_j}$, the corresponding total displacement noise spectrum becomes
\begin{equation}
S_{x}=\left|\eta\right|^{2}\sum_{j}\left|\chi_{\text{m,}j}\right|^{2}S_{F_{j}}.
\end{equation}

Within the context of sensing an externally applied force $\tilde{F}$ the corresponding noise floor is then
\begin{equation}\label{eq:S_F}
S_{F}=S_{x}/\left|\chi_{\text{m}}\right|^{2},
\end{equation}
where $\chi_\text{m}$ is the total susceptibility to $\tilde{F}$ including all modes. Note $\chi_\text{m}$ depends heavily on where and how the force is applied, since this changes the inertial (i.e., effective) mass of each mode. As a simple example, when sensing an external force that is applied at the same location as the measurement (e.g., when sensing the radiation force itself), we can calculate $\chi_\text{m}$ by setting $\tilde{F}_j\rightarrow \tilde{F}$ in Eq.~\ref{eq:total-displacement-noise}, such that
\begin{equation}\label{eq:total-susceptibility}
\chi_\text{m}\rightarrow\eta\sum_j\chi_{\text{m,}j}.
\end{equation}

To illustrate the characteristic spectral features of BANC with a concrete example, we calculate $\S_x$ and $S_F$ with the flexible mirror in Fig.~\ref{fig:1}(a) modeled as an ideal string, normalizing units so that $m_j=1$, $\Omega_j=j$, and temperature $k_BT=1$. We also assume the cavity mode is centered on the string, such that only modes having odd $j$ -- those with antinodes at the cavity field -- participate. To help with visualization, we assume a constant $Q_j\Omega_j$ product with $Q_1=1000$ and thermal force noise $S_{F_j}=4j^{2}/\omega Q_j$
associated with structural damping \cite{saulson1990thermal}. The solid black curves in Fig.~\ref{fig:1}(c) and (d) show the resulting displacement and force noise spectra for back-action strength $\xi=-0.7$ (anti-spring), including the first $N=1,001$ modes. As expected, each resonance peak in Fig.~\ref{fig:1}(b) is shifted to lower frequency, with the fundamental shifted farther than the (stiffer) higher-order modes. 

There are several important quantitative features associated with BANC. First, the frequency shift associated with a given $\xi$ is quite different than that predicted for a single mode (blue dashed line, $N=1$), owing to the contributed susceptibility of the higher-order modes at low frequency. In this case, the discrepancy is significant for the fundamental mode, even though the next lowest mode has three times larger frequency. Moreover, if one \textit{ignores} the influence of the other modes and adjusts $G$ by 20\% so that the shifted $\Omega_1$ matches observations, the expected broadband noise level is far from accurate -- as is the value \textit{at} resonance (inset) -- even when summing all the ``background'' contributions from the same $N=1001$ modes (each modeled with its own optomechanical loop and the same $G$). Moreover, the mechanical damping (linewidth) of the shifted mode is comparatively broadened, owing to the non-vanishing imaginary component of the higher-order mode susceptibilities; note we did not include any laser cooling in this model.

BANC provides access to an important technical advantage as well, which is readily visible in the total force noise floor (solid black line) of Fig.~\ref{fig:1}(d): despite the shift in resonance frequency, the low-noise bands -- where the $S_F$ achieves a value at (or a little below) that of a single mode (blue dashed lines) -- remain at the \textit{bare} mechanical frequencies $\Omega_j$. Importantly, this is now where the (modified) susceptibility has no sharp features, eliminating artifacts and noise associated with resonance frequency drift during measurement. 

Note the peaks in between these low-noise bands arise from destructive interference (anti-resonances) in the structure's response to external drive.

\begin{figure}[h]
\includegraphics[width=\columnwidth]{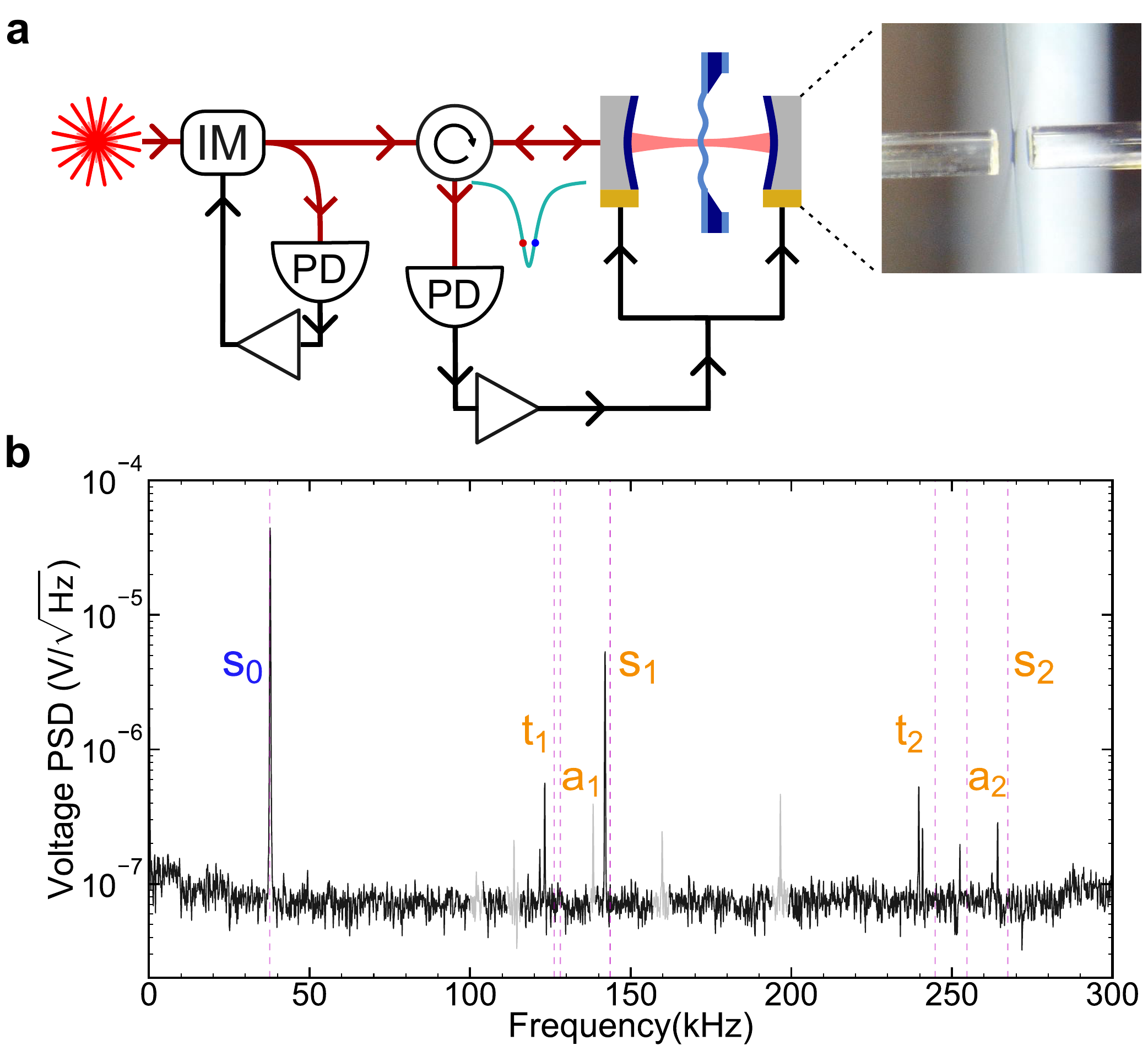}
\caption{\label{fig:2} 
Membrane-cavity system and measurement setup.
 (a) Optomechanical setup. Up to 2~mW of 1550~nm light is stabilized to near the shot noise limit by intensity modulator (IM) feedback \cite{dumont2023high} and / or 300~$\upmu$W of 1310 nm light drive an 80-\um-long fiber cavity with a \SiN ``trampoline'' \cite{reinhardt2016ultralow,norte2016mechanical} aligned near the center (inset image). The mirror coatings achieve finesse $\sim$7,000 at 1550 nm, while the 1310 nm light, outside the high-reflectivity band of the mirrors, generates roughly sinusoidal (very low finesse) fringes with negligible back-action. Reflected light from either beam is collected by a photodiode (PD), and the 1550 nm light is fed back to sheer piezos under the fiber mirrors to stabilize $\bar{\Delta}$ during measurement. In this case, the feedback gain is reduced until the feedback bandwidth $<$300 Hz, such that this does not influence the noise measurements over the bands plotted in Fig.~\ref{fig:3}; for measurements with 1310 nm light, feedback is not necessary. (b) ``Bare'' mechanical spectrum (using 1310~nm light), showing the fundamental ``symmetric'' mode \m{s_0} at 37.7 kHz, along with several higher-order symmetric (torsional) modes s$_j$ (t$_j$) and antisymmetric modes a$_j$ \cite{reinhardt2016ultralow}. Magenta vertical lines show mode frequencies predicted by COMSOL, and gray peaks correspond to high-mass chip modes and electronic noise peaks that do not play a role.}
\end{figure}

\textit{Observed back-action induced noise correlations.---} BANC becomes unmistakable when resolving thermal noise over a broad band. To access this regime, we assemble the optomechanical system drawn in Fig.~\ref{fig:2}(a), comprising a $\sim$80-\um-long fiber cavity \cite{hunger2010fiber}, etched \cite{bernard2020monitored} to finesse $\sim$7,000 for 1550~nm light, with a ``trampoline" \cite{reinhardt2016ultralow,norte2016mechanical} mechanical resonator -- a 90-nm-thick \SiN membrane patterned into a 100-\um-wide pad suspended by 1.5~\um-wide tethers within a 3~mm wide window -- suspended at the center, as shown in Fig.~\ref{fig:2}(b, inset). This is mounted on a vibration isolation stage at ultrahigh vacuum ($3\times 10^{-8}$~mbar), and addressed by either 1550~nm light for high-finesse optomechanics or 1310~nm light (outside the mirror coatings' high-reflectivity band) for low-finesse readout with negligible back-action (as in Ref.~\cite{Clark2024Nov}). To eliminate classical amplitude noise from the 1550~nm light, we pick off 6~mW upstream and feed back to an intensity modulator (IM), generating nearly shot-noise-limited output below $\sim$ 100~kHz \cite{dumont2023high}. Light reflected from the cavity is collected for readout and used to feedback-stabilize the cavity length, locking to the side of the resonance. Figure \ref{fig:2}(b) shows the trampoline's thermal mechanical spectrum acquired using only the 1310~nm laser. The labeled major peaks correspond to the ``symmetric'' ($\text{s}_j$), ``torsional" ($\text{t}_j$), and ``antisymmetric'' ($\text{a}_j$) modes, near predictions from COMSOL simulations (dashed lines). We perform ring-down measurements to determine the fundamental s$_1$ mode quality factor $Q_1 = (13.2 \pm 0.8) \times 10^6$. 

\begin{figure}[h!]
\includegraphics[width=\columnwidth]{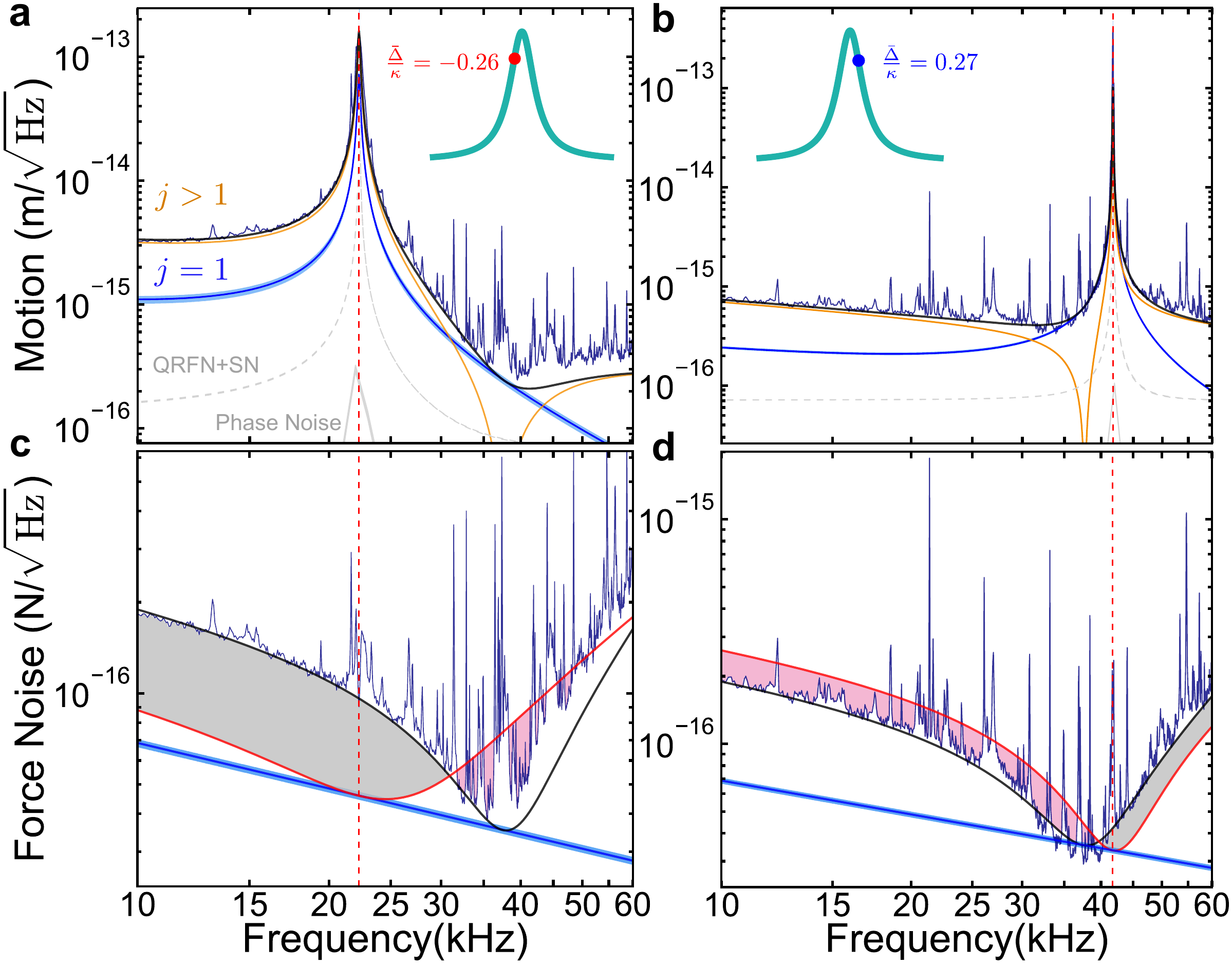}
\caption{\label{fig:3} Displacement (a)-(b) and force (c)-(d) spectra measured (dark blue) with the cavity driven by 1550~nm laser at detuning (incident power) $\bar{\Delta}=-0.26\kappa$ (22 $\upmu$W) and $\bar{\Delta}=0.27\kappa$ (12 $\upmu$W) for the left and right plots, respectively.  In (a) and (b), the black solid curve shows a fit to the BANC model using Eq.~\ref{eq:S_x_th_approximate}, assuming bare mechanical frequency $\Omega_1/2\pi = 37.7$ kHz, quality factor 13.2 $\times10^6$, temperature \textit{T}=295 K, susceptibility tail $\chi_\text{R}=0.457\pm0.003$ as fixed, independently measured parameters (see main text), and background noise tail 
$A_\text{T}$ = $4.95\times 10^{-26}$ m$^2$, $\Omega_{\text{eff}}/2\pi$ = 22.3 kHz (41.8 kHz) and $\Gamma_{\text{eff}}/2\pi$ = 1145 Hz (49 Hz) as free parameters for red (blue) detuning. The light blue curve shows the contribution from the first mode ($j=1$), and the orange shows that of the higher-order modes ($j>1$). Shaded regions represent the uncertainties, dominated by those of the quality factor and input coupling rate $\kappa_{\text{ex}}$ calibrated from measured cavity and membrane parameters (see main text). Dashed (solid) gray lines show contributed shot noise and QRFN (measured laser phase noise). (c) and (d) present force noise spectra from Eq.~\ref{eq:S_F}, with $\chi_m$ estimated from the fit parameters. Red lines in (c)-(d) show force noise estimated by summing all modes independently (no correlations), as in Fig.~\ref{fig:1}(c). Pink (grey) shading indicates bands where correlations suppress (enhance) the noise relative to this naive expectation. Single-mode sensitivity is approached at the bare mechanical frequency, despite thermal intermodulation noise (TIN) peaks, rather than at the resonance peak (red dashed line).}
\end{figure}

Figure \ref{fig:3}(a) and (b) show broadband motional spectra measured (dark blue) with 1550~nm light at detuning (input power) $\bar{\Delta}=-0.26\kappa$ (22 $\upmu$W) and $\bar{\Delta}=0.27\kappa$ (12 $\upmu$W), respectively. To calibrate the y-axis, we note that the amplitude quadrature of the reflected light (in the ``fast-cavity'' limit $\kappa\gg\Omega_j$) takes on the form  \cite{aspelmeyer2014cavity}
\begin{align}
    \tilde{X}_\text{out}=-\frac{G\sqrt{2\kappa_\text{ex}\bar n_\text{cav}} (\kappa-\kappa_\text{ex})\bar{\Delta}}{(\bar\Delta^2+\kappa^2/4)\sqrt{\bar{\Delta}^{2}+(\kappa/2-\kappa_\text{ex})^{2}}}\tilde{x},
\end{align}
where $\tilde{x}$ is the trampoline displacement, $\bar{\Delta}$ is the mean detuning, and $\kappa_\text{ex}$ is the cavity light's decay rate into the collection path. For the red (blue) detuning, the cavity linewidth $\kappa/2\pi=0.72$~GHz (0.68~GHz) (varies with trampoline position \cite{sankey2010strong,reinhardt2016ultralow}) and $\kappa_\text{ex} = $ 0.1$\kappa$ (0.18$\kappa$) estimated from the asymmetry of the reflected resonance lineshape \cite{janitz2015fabry}. The cavity photon occupancy $\bar n_\text{cav}=\kappa_{\text{ex}}\dot n_\text{in}/(\kappa^{2}/4+\bar\Delta^{2})$ for input photon rate $\dot n_\text{in}$ \cite{aspelmeyer2014cavity}, and the optomechanical coupling $G$ is estimated from the optical spring shift at low power \cite{aspelmeyer2014cavity}, where the corrections discussed above are negligible.

To model the thermal spectrum $S_x$ near and below the fundamental mode frequency, we approximate the total susceptibility $\chi_\text{m} \approx \chi_\text{m,1}+\chi_\text{R} + i\chi_\text{I}$, where the combined bare susceptibility of the higher-order modes (far below their own resonant frequencies) is approximated by a constant ``Hooke's law tail'' $\chi_\text{R}$ and non-vanishing imaginary component $\chi_\text{I}$ associated with structural damping at low frequencies \cite{saulson1990thermal}. Assuming $\chi_\text{I}\ll\chi_\text{R}$ (i.e., assuming high-$Q$ resonances), we can then approximate the back-action-modified susceptibility of the fundamental mode 

\begin{align}
    |\eta\chi_{\text{m,1}}| &\approx\left|\frac{\left(1+\xi\chi_\text{R}\right)^{2}}{1+\xi\chi_\text{R}-i\xi\chi_\text{I}}\chi_{1}^{-1}+\xi\right|^{-1} \\
    &\approx \left|\frac{\left[m_{1}(1+\xi\chi_\text{R})\right]^{-1}}{\Omega_{\text{eff}}^{2}-\omega^{2}-i\omega\Gamma_{\text{eff}}}\right|, 
\end{align}
 where $\Gamma_\text{eff}$ is the modified damping (including $\chi_\text{I}$, laser cooling, and/or photothermal effects \cite{clark2025detuning}), and
\begin{equation}\label{eq:effective-frequency}
\Omega_{\text{eff}}\approx\sqrt{\Omega_{1}^{2}+\frac{\xi/m_1}{1+\xi\chi_\text{R}}}
\end{equation}
is the effective (shifted) frequency. The optomechanical ``strength'' parameter $\xi\propto \mathcal{D}\propto n_\text{cav}\propto \dot n_\text{in}$ (input power), meaning the observed nonlinear dependence of $\Omega_\text{eff}$ on $n_\text{cav}$ allows an estimate $\chi_\text{R}=0.457 \pm 0.003$ (using $m_1=4.26$~ng estimated from COMSOL \cite{reinhardt2016ultralow}). Finally, assuming structural damping \cite{saulson1990thermal}, this approximation predicts a thermal displacement spectrum
\begin{align}\label{eq:S_x_th_approximate}
    S_x^{\text{Th}} &= |\eta\chi_\text{m,1}|^2S_{F_1}+  \sum_{j>1}|\eta\chi_{\text{m},j}|^2S_{F_j} \\
    &\approx |\eta\chi_\text{m,1}|^2\left(\frac{4m_1k_BT\Omega_1^2}{Q_1\omega}+\frac{1}{|\chi_{\text{m},1}|^2}\frac{A_\text{T}}{\omega}\right),
\end{align}
where the combined low-frequency noise from all of the higher-order modes ($\propto 1/\omega$) is captured by the phenomenological ``background noise tail'' parameter $A_\text{T}$. Note all noise sources are multiplied by $|\eta\chi_\text{m,1}|^2$ and therefore contribute significantly to the resonance peak at $\Omega_\text{eff}$, while the background term is \textit{divided} by $|\chi_\text{m,1}|^2$, suppressing its contribution near the original (bare) mechanical frequency $\Omega_1$. 

The presence of thermal intermodulation noise (TIN) \cite{fedorov2020thermal,Pluchar2023Nov} peaks precludes reliable least-squares fitting of the entire data set, but we can still use ``clean'' frequency bands to constrain $A_\text{T}$. The black line in Fig.~\ref{fig:3}(a)-(b) shows $S_x^\text{Th}$ assuming $T=295$ K, with background $A_\text{T} = 4.95 \pm 0.01 \times10^{-26}$~m$^2$, estimated by simultaneously fitting the ``clean'' background data below 20~kHz of both data sets, using effective frequency $\Omega_{\text{eff}}/2\pi = $ 22282 $\pm$ 3.8 (41761 $\pm 1.4$) Hz and damping rate $\Gamma_{\text{eff}}/2\pi=$ 1145 $\pm$ 8.5 (49 $\pm$ 2.7) Hz for red (blue) detuning constrained by data near the resonance peak.
The model then extrapolates to follow the noise floor (in between TIN peaks) over the majority of the measurement bandwidth for both detunings, capturing the BANC-induced enhancement and suppression of broadband noise at high and low frequencies expected for each situation. The deviations in Fig.~\ref{fig:3}(a) above 35 kHz are consistent with the broadband ``TIN forest'' observed with sufficiently low readout noise \cite{fedorov2020thermal,Pluchar2023Nov}. Also as expected, the orange curves in (a) and (b), show the background noise tail contribution (second term in Eq.~\ref{eq:S_x_th_approximate}), which is peaked at $\Omega_\text{eff}$ and suppressed at the bare mechanical frequency, where the observed noise approaches that of a single mode ($j=1$, light blue curves). 

To ensure these spectra are not driven by laser noise, the gray curves show the readout noise from our laser's amplitude (dashed) and phase (solid) noise, both of which are at least a factor of 10 smaller than observations. 

Under the same approximations, we can also estimate a noise floor for force measurements (at the measurement location) by dividing all of the curves in (a) and (b) by the magnitude of the \textit{total} susceptibility 
\begin{align}
    \chi_\text{m}\approx\frac{\chi_{\text{m,}1}+\chi_\text{R}}{1+\xi\left(\chi_{\text{m,}1}+\chi_\text{R}\right)}.
\end{align}
For reference, we have included a red curve representing the expectations from the aforementioned ``naive'' model (where one sums the noise of each mode individually, ignoring correlations, as in Fig.~\ref{fig:1}). The shading emphasizes the discrepancy, with pink highlighting broad regions where correlations produce noise observed \textit{below} these expectations, despite the TIN background. Finally, the red dashed lines highlight the observed fundamental mode resonance peak, where excess noise occurs primarily due to $\Omega_\text{eff}$ fluctuations during measurement. Importantly, this sharp feature is far from the band in which the lowest noise is observed.

\textit{Conclusions---}
We demonstrate displacement measurements capable of resolving broadband back-action-induced noise correlations in the thermal noise from many modes of a ``trampoline'' resonator. Importantly, the effect can be strong even if the modes are well-separated in frequency. Moreover, since the (shifted) mechanical resonance noise peaks are also strongly impacted, it is important to consider these effects in any system exhibiting such back-action effects, even if the readout noise is too high to resolve the broadband features. We also find that the lowest noise bands occur near the bare mechanical frequencies, even if the optical spring has shifted the resonance frequencies far away, and that this low-noise band approaches that of a single mode. 
Systems less susceptible to TIN, and/or those employing active TIN suppression (e.g., laser cooling \cite{aspelmeyer2014cavity}, multimode servos \cite{clark2025detuning}, or nonlinear noise canceling \cite{Huang2024Feb}), can in principle eliminate artifacts associated with resonance frequency noise. These low-noise bands should particularly benefit broadband measurements of quantum radiation force noise (QRFN)\cite{Cripe2019Apr} and ponderomotive squeezing \cite{Aggarwal2020Jul}, since the maximum visibility and degree of squeezing occur where thermal force noise is minimized. 
Finally, we note that these multimode interactions can actually suppress the force noise floor slightly \textit{below} that of a single-mode in these bands, due to the increased mechanical compliance afforded by the ``Hooke's law tail'' from the combined susceptilibity of the higher-order modes. This effect is too small to conclusively observe here, but is increasingly relevant when the external force is applied over a smaller area (i.e., exciting a broader band of higher-order modes).

\providecommand{\noopsort}[1]{}\providecommand{\singleletter}[1]{#1}%
%

\end{document}